# Bridging the gap between target-based and cell-based drug discovery with a graph generative multi-task model

Fan Hu[#], Dongqi Wang[#], Huazhen Huang, Yishen Hu and Peng Yin[*]

Guangdong-Hong Kong-Macao Joint Laboratory of Human-Machine Intelligence-Synergy Systems, Shenzhen Institute of Advanced Technology, Chinese Academy of Sciences, Shenzhen, 518055, China

#: These authors contributed equally to this work.

*To whom correspondence should be addressed: peng.yin@siat.ac.cn

## Abstract

Drug discovery is vitally important for protecting human against disease. Target-based screening is one of the most popular methods to develop new drugs in the past several decades. This method efficiently screens candidate drugs inhibiting target protein in vitro, but it often fails due to inadequate activity of the selected drugs in vivo. Accurate computational methods are needed to bridge this gap. Here, we propose a novel graph **m**ulti t**a**sk deep learning model to identify compounds carrying both **t**arget **i**nhibitory and **c**ell active (MATIC) properties. On a carefully curated SARS-CoV-2 dataset, the proposed MATIC model shows advantages comparing with traditional method in screening effective compounds in vivo. Next, we explored the model interpretability and found that the learned features for target inhibition (in vitro) or cell active (in vivo) tasks are different with molecular property correlations and atom functional attentions. Based on these findings, we utilized a monte carlo based reinforcement learning generative model to generate novel multi-property compounds with both in vitro and in vivo efficacy, thus bridging the gap between target-based and cell-based drug discovery.

# Introduction

Drug discovery is important for protecting human against disease. For the past several decades, target-based screening was one of the most popular methods to develop new drugs. This method is based on the hypothesis that a specific protein may have an important role in disease, thus disease can be treated by identifying a drug to inhibit the target protein. Target-based method first screens small molecules that can inhibit target proteins in vitro (e.g., enzymological experiment), and then verifies the effectiveness of them in vivo (e.g., cells experiment). Ideally, the in vitro inhibitors could penetrate cell membrane and inhibit target protein in vivo, thus repairing the biological functions destroyed by disease. One main advantage of target-based screening is that understanding the mechanism of action of drugs may facilitate yielding desired drugs without causing unexpected side effects. However, in the down sides, this method often fails due to inadequate activity of the selected drugs in vivo.

The high attrition rate of targeted drugs in vivo can be attributed to many possible reasons including inadequate drug exposure or different environments between in vitro and in vivo[1,2]. Inadequate drug exposure is often caused by inappropriate molecular properties of drugs. Mateus et al. proposed a label-free cellular method for quantifying the intracellular bioavailability of drug, and hence evaluated drug access to intracellular targets and its pharmacological effect[1]. Computationally, ADME (absorption, distribution, metabolism, and excretion) analysis is commonly utilized to select drugs with desired properties to increase cellular drug exposure. But this method may not ensure drug efficacy in vivo (as evaluated in Fig. 2A). Another cause is that the discrepancy between purified target enzyme in vitro and natural target enzyme in vivo [2]. Protein structure may vary depending on the environment in vitro and in vivo, thus altering the binding affinity between drugs and target protein. On the other hand, phenotypic-based screening could select effective compounds in cells directly, but it always suffers from target deconvolution and unexpected side effects [3]. Thus, considering the complexity within the process, it is necessary to develop a computational method to perform target-based and phenotypic-based screenings simultaneously, identifying compounds that are both effective in vitro (target inhibition) and in vivo (cell active) at the early stage of drug development.

In practice, many computational methods including AI (Artificial Intelligence) models have been developed to predict drug-target interaction, thus increasing the hit rates of target-based screening[4–6]. However, computational methods for accurate prediction of drug efficacy in vivo have been lacking. From a biochemical point of view, these two tasks are closely related. Therefore, multi-task learning model may open an avenue for solving this problem. Multi-task learning is an efficient strategy to learn multiple related tasks, facilitating knowledge transfer across these tasks[7–10]. For example, hard parameter sharing is a classical and common model but it may suffer from negative transfer due to the uncertain relationships between tasks.

In the present study, we propose a novel framework to bridge the gap between target-based and cell-based drug discovery, by taking drug development against SARS-CoV-2 as an application. As shown in Fig.1, our framework involves two parts: predictor and generator. First, a graph multi-task learning model

(MATIC) was proposed to predict both SARS-CoV-2 3CL$^{pro}$ inhibition in vitro and antiviral effect in vivo of compounds simultaneously. Next, based on this MATIC model, a monte carlo based reinforcement learning model was proposed to generate novel multi-property compounds with both target inhibitory and cell active properties.

The main advantage of our framework is that hidden information within each task could be shared and exchanged during training. For example, some compounds showed only in vitro or in vivo efficacy are probably effective and crossed in both tasks. We trained and evaluated our MATIC model on a carefully curated SARS-CoV-2 dataset derived from recent public databases and publications. We also tested the model on independent datasets[11–13] after removing duplicates from training set. Then, we explored the model interpretability and found that the learned features for target inhibition (in vitro) or antiviral (in vivo) tasks have different correlations with molecular properties. The visualization of the atom attention also showed that target inhibition and antiviral tasks focused on different functional groups of compounds. Based on these findings, we utilized the generator to produce novel multi-property compounds with dual-v (in vitro and in vivo) efficacy combined with MATIC, thus bridging the gap between target-based and cell-based drug discovery.

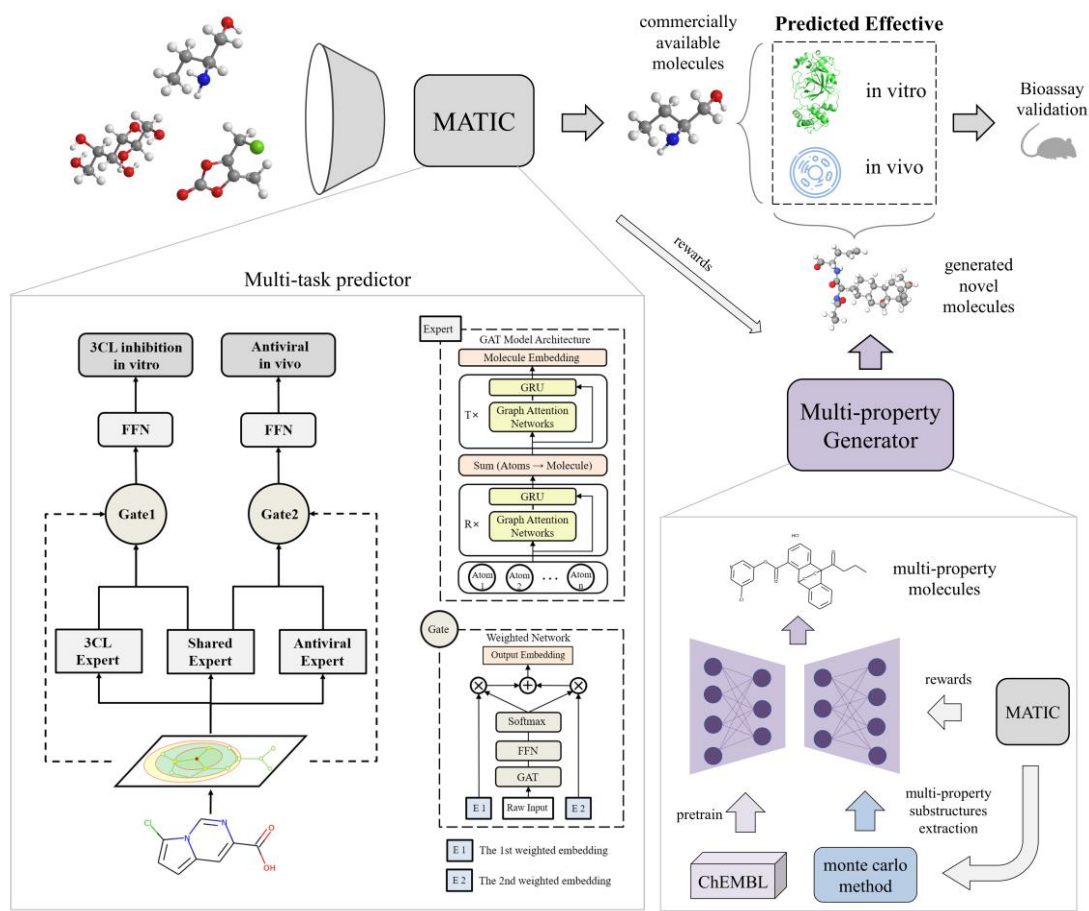

Fig. 1 Schematic of the proposed framework. The framework involves two parts: predictor and generator. MATIC (multi-task predictor) consists of three experts and two gates and predicts molecules that are effective in vitro and in vivo. The generator produces novel multi-property (effective in vitro

and in vivo) molecules after pretrained by ChEMBL dataset and finetuned by mento carlo based method and MATIC.

## Results and Discussion

Data collection and analysis

The SARS-CoV-2 3CL$^{pro}$ and antiviral data were collected from various papers [14–20] and public datasets such as PubChem, as listed in Supplementary Table 1. After removing duplicates, we obtained a collection of total 7,458 compounds. Most of these compounds have validated $IC_{50}$ or $EC_{50}$ values and we chose 20 μM as the threshold to determine the positive and negative samples. The value of 20 μM was commonly used in biochemical screening as a cutoff to categorize active and inactive compounds [2,20]. Finally, there are 356 compounds with both 3CL and antiviral labels, in which at least one label is positive. There are 2,729 compounds with either 3CL or antiviral label. The rest 4,373 compounds are negative for both 3CL and antiviral. Next, we calculated multiple molecular properties such as lipophilicity and physicochemical properties of these compounds using SwissADME[21].

Traditionally, ADME analysis was used to select drugs with desired properties to increase cellular drug exposure. To evaluate the efficiency of this method, we simulated a screening process to select effective compounds in vivo (antiviral) from targeted compounds (3CL inhibition) by traditional manual rules filter according to molecular properties [22,23] (Fig. 2A). Specifically, for the collected dataset, there are 4463 compounds with both 3CL and antiviral labels. Among them, there are 269 compounds that can inhibit 3CL in vitro. Then, we selected 167 of 269 compounds according to traditional filters: $150 \leq$ molecular weight $\leq 500$, $-0.7 \leq LogP \leq 5$, $-6 \leq LogS \leq 0$, $TPSA \leq 140$. For the selected 167 compounds, only 38 of them were positive in vivo while 129 were negative. Whereas the attrited 102 compounds had 41 positive and 61 compounds. The AUC score of this traditional method was 0.4, suggesting that the traditional methods based on molecular properties may not efficiently screen effective compounds in vivo.

Furthermore, we drew the distributions of these properties for compounds with 3CL or antiviral positive labels, to clarify if there remains a clear gap between these compounds. As shown in Fig. 2C, the numbers of 3CL inhibitors and antiviral compounds are approximately equal while the both positive compounds are only a small proportion. For the exhibited four molecular properties, the distributions of 3CL inhibitors and antiviral compounds are mostly overlapping. More specifically, the LogP values and the synthetic accessibility scores of antiviral compounds are slightly larger, suggesting the higher permeability and more difficult to synthesis. Whereas the LogS values of 3CL inhibitors are slightly larger, indicating the higher solubility in water. In summary, it may not be straightforward to distinguish the 3CL and antiviral compounds only based on these molecular properties.

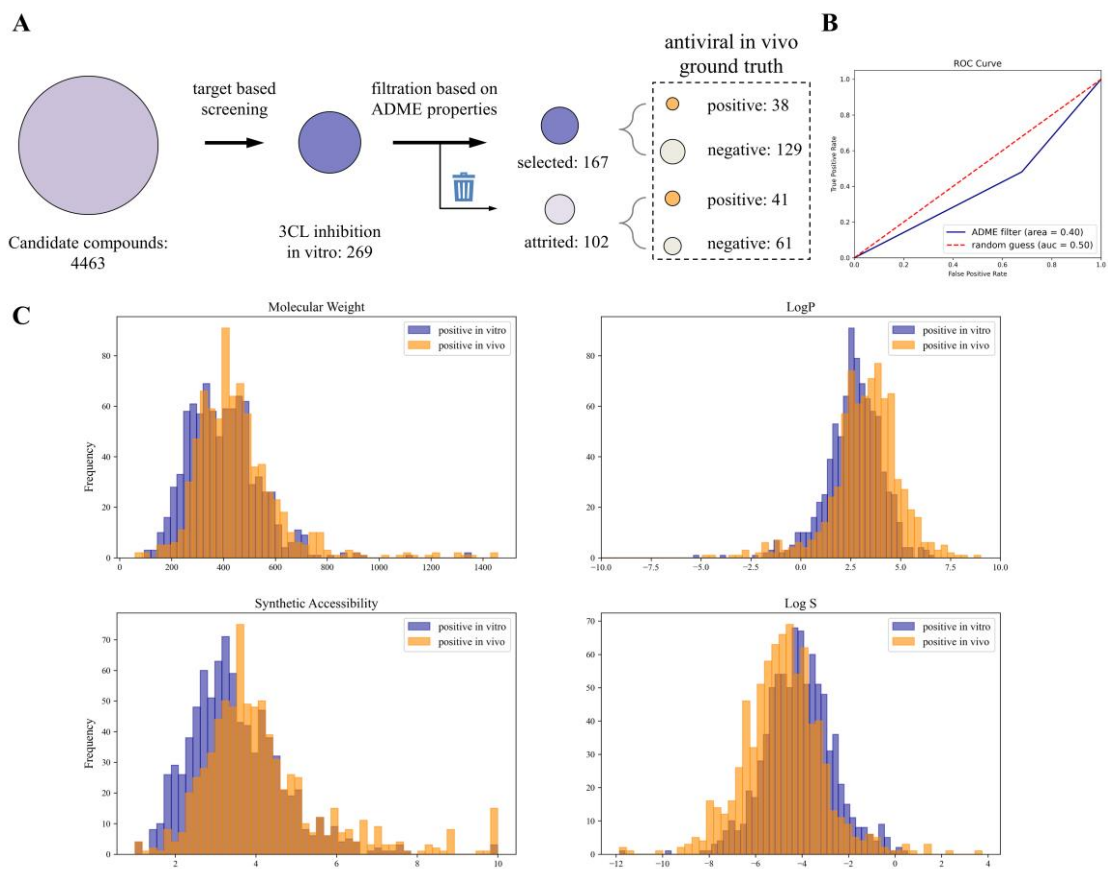

Fig. 2 The analysis of the collected dataset and related ADME properties. (**A**) Simulated screening of effective compounds in vivo based on ADME properties filter. (**B**) The ROC curve of ADME filter. (**C**) Distribution of molecular properties of compounds that are effective in vitro or in vivo. Molecular properties are as follows. Molecular weight, the mass of a given molecule (unit - Dalton). LogP (partition coefficient/lipophilicity), an increase in LogP enhances permeability. LogS (water solubility values), the higher the LogS value, the higher solubility of the compound. Synthetic Accessibility, the score is ranging from 1 (easy synthesis) to 10 (very difficult synthesis).

Model training and performance

Initially, we desired to select a graph-based neural network for processing molecule as the base module of our multi-task model and also to acquire baseline metrics of single-tasks (i.e., 3CL inhibition in vitro and antiviral in vivo) on the curated SARS-CoV-2 related dataset, as described in methods section. For training and evaluation, compounds with at least one positive label were selected and double negative compounds were chosen according to the same proportion to avoid an imbalance of data. We trained and evaluated GAT and GROVE [25] for single-tasks on our data using 3-fold cross-validation. It should be noted that GAT was trained from scratch whereas GROVE was pretrained on a large-scale molecular dataset (GROVE$_{base}$ model and weights used in their paper) and then fine-tuned on our dataset. The results show that GAT achieved better performance than GROVE (Table 2). Probably, the knowledge learned by pretraining may not be relevant to this task and thus resulting in negative transfer. Another possible explanation is that GROVE with too much parameters may be overfitting on a relatively small dataset. Therefore, GAT was utilized as an expert module of multi-task not only the better performance but also its lighted-weight. Another advantage is that atom weights processed by GAT could be easily visualized and thus providing a biological interpretation.

Then we compared multi-task models MMOE and our proposed MATIC. MMOE model utilized three shared experts (GAT), two specific gates and two fully connected layers to predict two tasks separately. For the proposed MATIC model, as shown in Fig. 1, gate1 gathered information from 3CL expert, shared expert and raw molecule input and assigned weights to get a molecule representation vector for 3CL inhibition task. Gate2 similarly produced a molecule representation vector for antiviral task. The Adam optimizer was used to train these models and the learning rates, batch size and dropout were set to 0.0015, 64 and 0.2, respectively.

Table 1. Model performance comparisons on 3CL$^{pro}$ and antiviral dataset.

| Model | Task type | AUC | Acc | Precision | Recall | F1-score |
|---|---|---|---|---|---|---|
| Single-task (GROVE) | 3CL | 0.83 | 0.753 | 0.7 | 0.667 | 0.683 |
|  | Antiviral | 0.743 | 0.67 | 0.783 | 0.35 | 0.484 |
| Single-task (GAT) | 3CL | 0.858 | 0.777 | 0.749 | 0.67 | 0.707 |
|  | Antiviral | 0.813 | 0.749 | **0.803** | 0.557 | 0.658 |
| Multi-task (MMOE/GAT) | 3CL | 0.88 | 0.79 | **0.752** | 0.711 | 0.731 |
|  | Antiviral | 0.82 | 0.753 | 0.782 | 0.607 | 0.683 |
| Multi-task (MATIC) | 3CL | **0.88** | **0.795** | 0.741 | **0.747** | **0.744** |
|  | Antiviral | **0.832** | **0.761** | 0.784 | **0.625** | **0.696** |

Table 1 exhibits average results over cross-validation folds for each method on our curated dataset, as shown, our proposed MATIC has achieved the best performance on most metrics. The high recall value indicated that MATIC has found more true positive compounds. This can be attributed to that the related information from different tasks could be better shared and used, even without corresponding labels. For example, many compounds showed only 3CL inhibition but actually had antiviral activity cannot be used for training in antiviral single-task because of label missing, but they are probably effective and crossed in both tasks of multi-task model. Given this characteristic, MATIC could not be only used for identifying both 3CL inhibitory and cell active compounds, but also be utilized for target deconvolution. Additionally, MATIC has achieved better performance than MMOE, suggesting that separating specific and shared experts can ensure knowledge transfer across tasks and guarantee the completeness of task specific features. To further evaluate the generalizability, we also tested MATIC on three independent sets including 3CL set 1[11], 3CL set 2[26] and antiviral set[13] after removing duplicates from the whole set (described in detail in Supplementary Table 2).

## The gap between target inhibitory and cell active compounds

As exhibited in previous studies [2,18], a general correlation between $3CL^{pro}$ inhibition and antiviral effect was not found. That is, the high inhibitory activity for $3CL^{pro}$ of a compound cannot guarantee its antiviral effect. Actually, there remains a gap between target inhibitory and cell active of compounds during drug development due to various reasons such as cell permeability and metabolic stability. Traditional ADME analysis based on only molecular properties may not efficiently pick out in vivo effective compounds (Fig. 3A). It is difficult to calculate an index that quantifies such gap based on these factors. Although our model has achieved excellent performance on predicting both 3CL and antiviral compounds, the results from a black box model may raise the risk of following false leads and it is hard to optimize those compounds with either high 3CL or antiviral activity. Therefore, we explored the mapping of learned features by our model to molecular properties and the corresponding important sites for 3CL and antiviral tasks discerned by our model.

**Mapping of learned features to molecular properties.** As mentioned above, we collected multiple molecular properties of compounds in our dataset from SwissADME [21], including lipophilicity, water solubility. Then we calculated the Pearson correlation coefficient between each dimension of the learned 300-dimensional feature vectors of molecules and the molecular properties. As shown in Fig. 3B, for molecular weight and synthetic accessibility, the correlation distributions of antiviral task have changed more after training compared to that of 3CL task. The higher positive correlations indicate that antiviral task may focus more on large or complex compounds. For LogS value, antiviral task showed higher negative correlations than 3CL task. As mentioned, the higher the LogS value, the higher solubility of compound. It seems that antiviral task prefers lower water solubility of a compound and it makes sense that high water solubility is not conducive to cellular drug exposure. However, the correlations with LogP value of both tasks did not changed significantly after training. LogP value correlates positively with cell permeability. One possible explanation is that most compounds in training set may have suitable permeability that model did not take it as an important feature to meet task requirement. Combined with the above results, it may not be easy to select in vivo effective compounds manually based on one or some related properties, but the MATIC model has captured some hidden correlations across molecular properties, thus distinguishing in vitro and in vivo tasks.

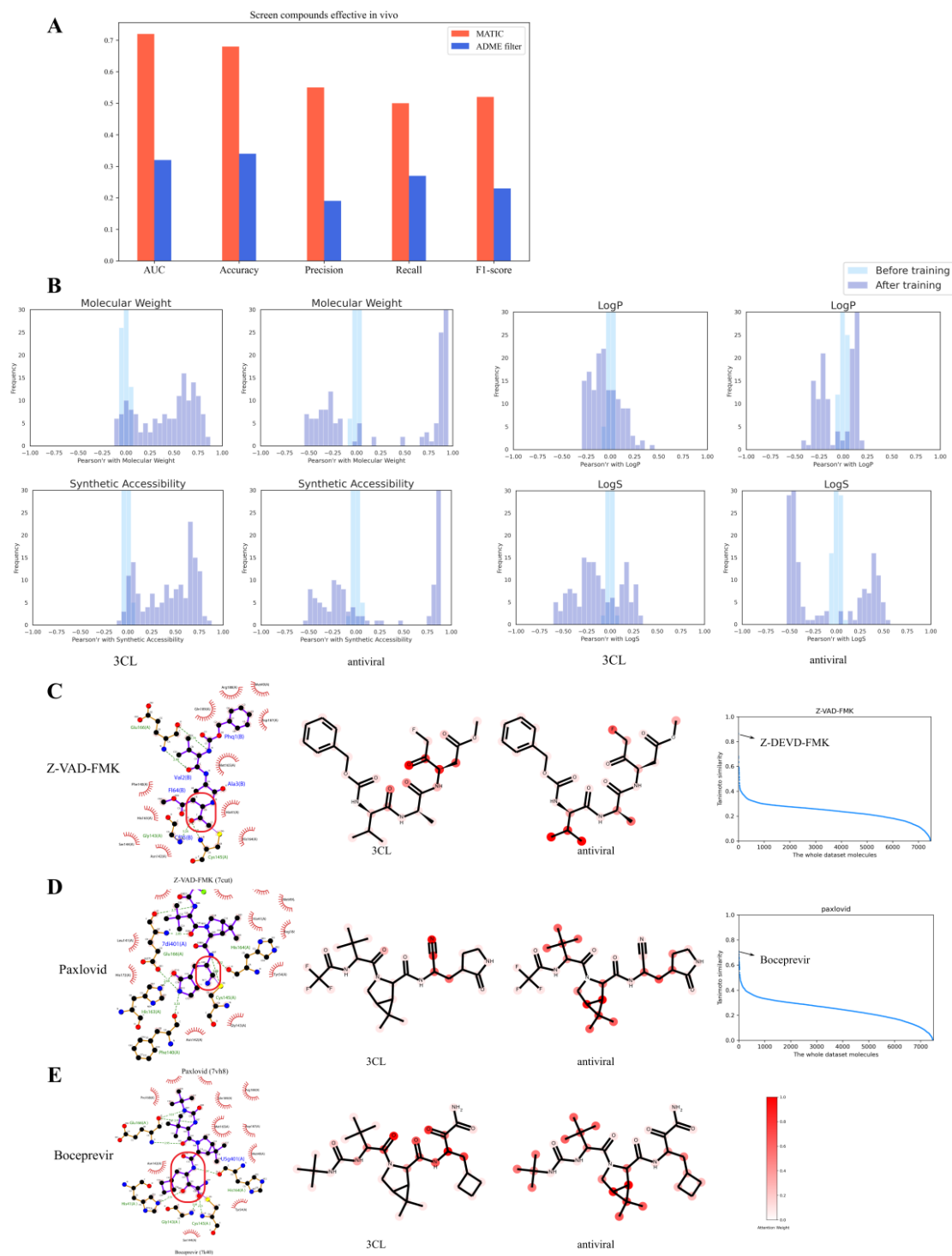

Fig. 3 The gap between target inhibitory and cell active compounds. (**A**) Performance comparison of simulated screening of effective compounds in vivo. (**B**) Mapping of learned features to eight molecular properties including MW, LogP, LogS and synthetic accessibility. The left picture is for 3CL task of each subplot and the right is for antiviral task. (**C-E**) Left: the interactions between Z-VAD-FMK (**C**), paxlovid (**D**), boceprevir (**E**) and 3CL$^{pro}$ (PDB ID: 7cut, 7vh8, 7k40). Middle: the predicted important atoms of Z-VAD-FMK (**C**), paxlovid (**D**), boceprevir (**E**) for 3CL$^{pro}$ (left) and antiviral task

(right). Right: the Tanimoto similarity scores between Z-VAD-FMK (**C**), paxlovid (**D**) and our whole set and the arrow indicates the most similar compound.

**Visualization of atom attention.** The final representation vector of a molecule for either 3CL or antiviral task is a weighted combination from gates, task-specific experts and shared expert. It raises the question of whether the same atoms or functional groups of a compound contributing to both 3CL and antiviral tasks? As the molecule representation is formulated by its atoms and bonds embedding vectors through the attention mechanism, we visualized the key atoms ranked by attention weight of molecule representation for different tasks.

To evaluate the model performance on predicting key atoms, we selected several important compounds that have crystal structures in complex with SARS-CoV-2 3CL$^{pro}$ including GC376 and MI-23 (discussed in detail in Supplementary Figure S1). We further chose another two compounds, Z-VAD-FMK from independent test set 1 and paxlovid (a recent reported effective drug in clinical trial for COVID-19) to test the generalizability of our model. Recent studies have reported that Z-VAD-FMK was an effective covalent inhibitor for SARS-CoV-2 3CL$^{pro}$ [11,27,28]. The C-terminal warhead (fluoromethyl ketone) of Z-VAD-FMK could stable covalently bind to the residue Cys145 of 3CL$^{pro}$ by a nucleophilic attack. In our prediction, the 3CL inhibition of Z-VAD-FMK was predicted as positive whereas the antiviral was predicted as negative which were consistent with the ground truth. The warhead was clearly highlighted for 3CL task and the side chains of valine and alanine were captured for antiviral task (Fig. 3C). These two amino acids valine and alanine are hydrophobic, corresponding to cell permeability. Moreover, the Tanimoto similarity scores [29] (0-1, the more closer to 1 the more similar it is) between the two compounds and each compound of the whole dataset including the training, validation and test sets are showed in Fig. 3C, D. The Tanimoto nearest neighbor in the whole dataset of Z-VAD-FMK and paxlovid were Z-DEVD-FMK (0.86) and boceprevir (0.71), respectively. The true double positive paxlovid were confidently predicted as double positive by our model. The nitrile carbon, which formed a covalent bond with Cys145 and its surrounded atoms have been highly weighted by our model for 3CL task. More interestingly, the key regions of paxlovid's nearest neighbor boceprevir that formed the critical covalent bond with Cys145, although are mostly different from that of paxlovid, have also been captured by our model. These results suggest that our model does not only rely on the local structural similarity but also probably relies on global message passing, and thus showing a good generalizability.

Bridging the gap by multi-property molecular optimization

As observed above, our MATIC model has captured some important features that contribute to 3CL and antiviral tasks. Interestingly, different tasks focus on different functional groups and the gap clearly remains. Based on this observation, we desired to extract substructures that were recognized important in 3CL or antiviral tasks by MATIC model, and then generate novel multi-property molecules (i.e., have both 3CL and antiviral) using these substructures. To bridge the gap, we explored different methods of multi-property molecular optimization. However, we found that reinforcement learning method that simply added multiple rewards may easily get stuck and the generated molecules may have similar simple structures, although a penalty of molecular diversity has been added. As shown in Fig. 4A, we extracted 1,000 single-property substructures from each task of 3CL and antiviral tasks according to distribution probability of MATIC and composed novel molecules by these substructures. Next, we selected 10,000 from these novel multi-property molecules based on high scores given by MATIC and completed them by a graph based variational autoencoder trained via reinforcement learning.

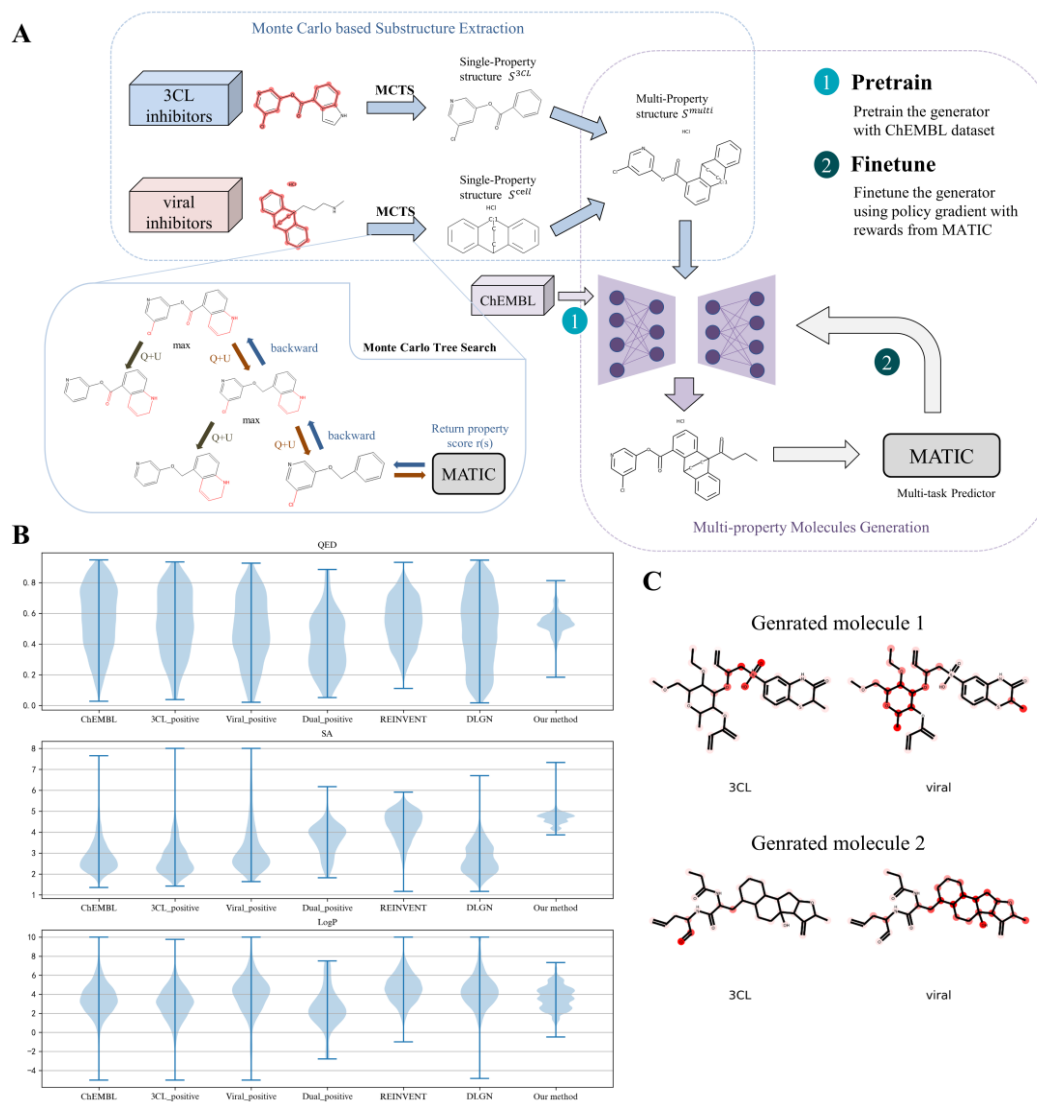

Fig. 4 Bridging the gap by multi-property molecular optimization. (**A**) Schematic of multi-property molecular optimization. (**B**) Distributions of QED, LogP and SA score for original and generated molecules. (**C**) The attention weights of selected generated molecules indicate important atoms for 3CL and antiviral tasks.

**Generator performance.** To fully evaluate model performance, we used several metrics including validity, success, novelty and diversity, which are widely used for comparison of molecule generative methods. Validity is the fraction of valid molecules among total generated molecules. Success is the fraction of positive molecules (i.e., have both 3CL and antiviral) among total generated molecules.

Diversity assesses the diversity of generated molecules by calculating the Tanimoto distance between generated and training molecules. The higher the diversity score, the better the molecular diversity. Novelty is defined as the fraction of generated molecules with their nearest neighbor similarity of the training positive molecules lower than 0.4 among total generated molecules. We compared our method to several popular multi-property generative methods. As shown in Supplementary Table 2, our method achieves 99% valid hit rate and 96% success rate while maintaining 0.97 novelty and 0.88 diversity, which is higher than other methods. It should be noted that most generated molecules produced by other methods consist of simple substructures, although these methods have achieved relatively high novelty and diversity scores. This may due to the difficulty in capturing syntax of complex structures, as well as reward sparsity.

**Molecular properties of generated molecules.** We further evaluated important molecular properties of the generated molecules, including QED, LogP and Synthetic Accessibility (SA). As mentioned above, LogP represents partition coefficient and an increase in LogP enhances permeability of molecules. QED (Quantitative Estimate of Druglikeness) score measures how likely a molecule can be a drug and QED score is in the range of 0 to 1, the higher score the better. SA score refers to how easy a molecule can be synthesis and is ranging from 1 (easy synthesis) to 10 (very difficult synthesis).

We sampled 10,000 molecules from each method and calculated their distributions of molecular properties. As shown in Figure 6, molecules generated by our method have more concentrated molecular properties. For example, QED scores of molecules generated by our method are mainly between 0.4 and 0.7, indicating most molecules have high drug-likeness. In contrast, molecules generated by other methods have wide distributions of QED scores, suggesting instability of model generation. For SA score, the scores of ChEMBL original molecules are distributed between 2 and 4 which indicates most molecules have simple structures. Whereas the relatively higher SA scores of molecules generated by our method suggest that these molecules have more complex structures which need to satisfy desired multiple properties.

**Visualization of generated molecules.** We visualized some generated molecules with high predicted scores. As shown in Fig. 4C, the substructure that contributes to 3CL task of compound 1 is obviously derived from GC376 (Supplementary Figure S1), which could form a covalent bond with Cys145 in the active sites of 3CL$^{pro}$ as aldehyde form. Compound 2 contains substructures similar to MI-23 (Supplementary Figure S1), which could form a covalent bond with sulfur atom of Cys145 by its carbon of the warhead aldehyde. On the other hand, substructures corresponding to antiviral task of both compound 1 and 2 have greatly improved the cell permeability. The LogP values of compound 1 and 2 are 2.37 and 3.05 while that of GC376 and MI-23 are -2.18 and 2.51. As mentioned above, the higher LogP value the better cell permeability of compound. Therefore, the generated compound 1 and 2 may have high 3CL inhibitory activity while maintaining considerable permeability which allow them to target 3CL within cell. We have exhibited more generated compound with attention weights in supplementary information.

Prediction of commercially available compounds

Considering that the generated compounds are not easily available, we also screened a commercially available library containing 500,000 lead-like compounds using the proposed MATIC model. There were 331 compounds with predicted probabilities more than 0.9 for both 3CL and antiviral. Then we selected a collection of 20 compounds with high molecular diversity (10 of them are listed in Supplementary Figure S2). Some key functional groups (e.g., nitrile carbon) that were highlighted in the above covalent inhibitors can be found in these selected compounds. We also visualized atom weights of these 20 compounds for both 3CL and antiviral tasks (Supplementary Figure S4). Moreover, the ADME properties of these compounds were also predicted (Supplementary File S2). Most of these compounds fell within the Lipinski's rule of five, suggesting high druglikeness.

## Conclusion

One main reason for the high failure rate of target-based drug discovery is that the selected effective compounds in vitro (target inhibition) may be ineffective in vivo (cell active). To address this challenge, we propose a framework to solve this problem and the main contributions are as follows. First, we have constructed a dataset related to SARS-CoV-2 which contains effective compounds in vitro and in vivo. Second, based on this dataset, we demonstrated that traditional method related to ADME properties may not be able to accurately select in vivo effective compounds. Third, we proposed a graph multi-task deep learning model, namely MATIC, to predict compounds that are both effective in vitro and in vivo. Last, we presented a reinforcement learning based dual-v (in vitro and in vivo) generative model to produce novel multi-property compounds, thus bridging the gap between target-based and cell-based drug discovery.

There are still some unresolved issues. For example, the theoretical basis for the underlying mechanism from in vitro effective to in vivo effective of compounds are obscure. Another important issue is that the protein families and networks should be necessarily modeled especially for complex disease. It should also be noted that molecular properties of compounds may not be the key points in effectivity evaluation. The compensatory mechanism among protein networks may activate another disease associate pathway to detour the inhibited targets. Our future study will further explore these challenges and extend in silico and biological findings into theories.

# Materials and Methods

Multi-task predictor (MATIC)

**Multi-task network architecture**

Basically, the model mainly consists of three parts, expert network, gating network and the tower network. The expert and gating networks accept molecule structure as input. The expert networks can be further divided into shared expert network and task specific expert network, which are responsible for, respectively, learning common knowledge across different tasks and specific domain knowledge from each task. Thus the molecule representation for k th task is formulated as:

$$S^k(x) = [E_k^T, E_s^T]^T \qquad (1)$$

where $E_k^T$ and $E_s^T$ are the outputs of task-specific and shared expert networks, respectively. Then both expert networks are fused and weighted by a gating network to get the final representation of molecule. The output of the gating network is calculated as:

$$w^k(x) = Softmax(g(x)) \qquad (2)$$

where $g(x)$ is the output of gating network. The final representation and prediction of task k are formulated as:

$$O^k(x) = w^k(x)S^k(x) \qquad (3)$$

$$y^k(x) = T^k(O^k(x)) \qquad (4)$$

$T^k$ is the tower network consisting of several fully connected layers.

The pseudocode of the proposed MATIC is provided in Algorithm 1.

---

**Algorithm 1** MATIC

1: **Input**: molecule x, the number of specific-task experts m, the number of shared experts n, all tasks K.
2: **Output**: predictions of all tasks
3: **for** each shared expert j=0 to n **do**
4:     $E_{sj} = e_{sj}(x)$
5: **for** each task k=1 to K **do**
6:     **for** each specific-task expert i=1 to m **do**
7:         $E_{ki} = e_{ki}(x)$
8:     $W_k(x) = Softmax(g(x))$
9:     $S_k(x) = [E_{k1}^T, E_{k2}^T, \ldots, E_{km}^T, E_{s1}^T, E_{s2}^T, \ldots, E_{sn}^T]^T$
10:    $O_k(x) = W_k(x)S_k(x)$

11:     $y_k(x) = T^k(O_k(x))$
12:   **Return** $y_1(x), y_2(x), ..., y_K(x)$

**Graph attention network**

Graph Attention Network (GAT) is used to extract molecular representation and fuse representations in our expert and gate networks, respectively. Here we used a GAT structure similar to Attentive FP [24]. More specifically, the representation vector of node i for one graph attentional layer is calculated as:

$$e_{ij} = a(W\vec{h}_i, W\vec{h}_j) \quad (5)$$

$$\alpha_{ij} = softmax(e_{ij}) = \frac{\exp(e_{ij})}{\sum_{k \in N_i} \exp(e_{ik})} \quad (6)$$

$$\vec{h}_i' = \sigma(\sum_{j \in N_i} \alpha_{ij} W\vec{h}_j) \quad (7)$$

where $\vec{h}_i$, $\vec{h}_j$ are the representation vectors of target node i and neighbor node j. All neighbors of node i are represented by $N_i$. The representation vector of node i $\vec{h}_i'$ can be obtained by aggregating information of every neighbor in the graph through attention. Then the GAT was processed for *l* iterations, similar to Graph Neural Network, involving message passing and readout phases:

$$\vec{h}_i'^{(l-1)} = \sum_{j \in N_i} M^{l-1}(\vec{h}_i^{(l-1)}, \vec{h}_j^{(l-1)}) \quad (8)$$

$$\vec{h}_i^l = GRU^{l-1}(\vec{h}_i'^{(l-1)}, \vec{h}_i^{(l-1)}) \quad (9)$$

where $M^{l-1}$ represent graph attention mechanism and GRU is activate function gated recurrent unit[32]. $\vec{h}_i^l$ represents node i after *l* iterations.

**Loss function**

The final loss is determined by the loss weights of the two tasks, which can be formulated as:

$$Loss = \sum_{k=1}^{k} w_k L_k \quad (10)$$

where $w_k$, and $L_k$ is the weight and loss of task k, respectively. Considering there are many masked label (label for either single task is unknown), the shared parameter part is updated for all samples and the specific parameter part is updated for sample with corresponding label.

Multi-property generator

We used a generative model which was similar to RationaleRL[33]. As shown in Fig. 4A, there were three main parts to generate novel compounds that are both target inhibitory and cell active. First, single-property substructures for either target inhibition or antiviral tasks were extracted using monte carlo algorithm based on MATIC model. Then, multi-property substructures were produced by combining these single-property substructures. Finally, novel compounds were completed by expanding multi-property substructures through a graph based variational autoencoder (VAE).

**Single-property substructure**

We called the substructure S which was associated with a particular property as single-property substructure, and S was a connected subgraph of full molecule G. Monte Carlo Tree Search (MCTS) was used to search single-property substructures for each molecule. Generally speaking, the root node of the search tree is a complete positive molecule G and other branch nodes of the search tree are chemically effective connected substructure which are obtained by deleting the peripheral bonds or peripheral rings from G. MCTS requires a large number of iterations, and each iteration consists of two steps. The first is the forward search phase. Search a path from the root node to the leaf node with less than N atoms. The search strategy used the PUCT algorithm[34] to calculate the score of each branch of the current state $s_k$ and then select the branch $a_k$ with the highest score:

$$U(s_k, a) = c_{puct} R(s_k, a) \frac{\sqrt{\sum_b N(s_k, b)}}{1 + N(s_k, a)} \quad (11)$$

$$Q(s_k, a) = \frac{W(s_k, a)}{N(s_k, a)} \quad (12)$$

$$a_k = \arg\max_a Q(s_k, a) + U(s_k, a) \quad (13)$$

where $(s_k, a)$ represents the branch obtained by deleting the peripheral bond or peripheral ring a from the state $s_k$. N($s_k$,a) is the number of visits. W($s_k$, a) is the total action value. Q($s_k$, a) is the average action value and R($s_k$, a) is the property prediction score of the substructure predicted by the MATIC model. Then there is the backward phase, calculating the property prediction score R of the leaf node and updating the information saved by the parent node.

$$N(s_k, a_k) = N(s_k, a_k) + 1 \quad (14)$$

$$W(s_k, a_k) = W(s_k, a_k) + R \quad (15)$$

After a large number of searches, the method will tend to search for substructures with higher scores. Finally, we selected the leaf nodes whose property score $R > \delta$ and the number of atoms less than N in the search tree. Then we constructed single-property substructures libraries $V^{3CL}$ and $V^{cell}$.

**Multi-property substructure**

Next, we used $V^{3CL}$ and $V^{cell}$ to produce multi-property substructures library. Briefly, we superposed $S^{3CL}$ and $S^{cell}$ to ensure their maximum common substructure concedes[1]. We then got candidate multi-property substructures library $C^{multi}$:

$$C^{multi} = \{MERGE(S^{3CL}, S^{cell}) \mid S^{3CL} \in V^{3CL}, \ S^{cell} \in V^{cell}\} \quad (16)$$

In addition, each $S^{multi} \in C^{multi}$ should satisfy property constrains.

$$V^{multi} = \{S^{multi} | S^{multi} \in C^{multi}, R^{3CL}(S^{multi}) > \sigma, R^{cell}(S^{multi}) > \sigma\} \quad (17)$$

Where $V^{multi}$ is the final multi-property substructures library and $R^{3CL}(S^{multi})$ and $R^{cell}(S^{multi})$ are property scores predicted by MATIC and $\sigma = 0.5$.

**Compound completion**

A graph based variational autoencoder trained via reinforcement learning completed a full molecule given a substructure from $V^{multi}$. To ensure the model generate valid compounds with desired properties, we trained the model in two steps. First, we pre-trained model on ChEMBL dataset to ensure model could generate valid compounds given random substructure. ChEMBL dataset was preprocessed into (S, G) and S is a random connected subgraph of the molecule G. Finally, we finetuned the model using policy gradient with reward from MATIC predictor.

## Evaluation metrics

Several metrics including precision, accuracy, recall and Matthews correlation coefficient (Mcc) was used to evaluate model performance on the 3CL and antiantiviral dataset:

$$Precision = \frac{TP}{TP+FP} \qquad (18)$$

$$Accuracy = \frac{TP+TN}{P+N} \qquad (19)$$

$$Recall = \frac{TP}{TP+FN} \qquad (20)$$

$$Mcc = \frac{TP \times TN - FP \times FN}{\sqrt{(TP+FP)(TP+FN)(TN+FP)(TN+FN)}} \qquad (21)$$

where TP, TN, FP and FN represent the number of true positives, the number of true negatives, the number of false positives and the number of false negatives, respectively. P and N indicate positive negative.

# Acknowledgements

This work was supported by the Strategic Priority Research Program of Chinese Academy of Sciences (NO. XDB 38040200) and the Shenzhen Science and Technology Innovation Committee (JCYJ20180703145002040).